\documentclass[aps,
superscriptaddress,reprint]{revtex4-1}
\usepackage{graphicx}
\usepackage{setspace}
\usepackage{amsmath}
\usepackage{amssymb}
\usepackage{color}
\usepackage{array}
\usepackage{subfigure}
\usepackage{hyperref}
\usepackage{float}
\usepackage{lipsum}
\usepackage{ulem}

\usepackage{multirow}

\usepackage[all]{xy}
\newcommand{\RN}[1]{%
  \textup{\uppercase\expandafter{\romannumeral#1}}%
}

\begin{document}
\title{Qubitization of Bosons}

\author{Xin--Yu Huang}
\affiliation{College of Physics, Mechanical and Electrical Engineering, Jishou University, Jishou 416000, P.R.China}

\author{Lang Yu}
\affiliation{College of Physics, Mechanical and Electrical Engineering, Jishou University, Jishou 416000, P.R.China}

\author{Xu Lu}
\affiliation{College of Physics, Mechanical and Electrical Engineering, Jishou University, Jishou 416000, P.R.China}

\author{Yin Yang}
\affiliation{College of Physics, Mechanical and Electrical Engineering, Jishou University, Jishou 416000, P.R.China}

\author{De--Sheng Li  }
\email[email: ]{lideshengjy@126.com. 
Xin-Yu Huang, Lang Yu, Xu Lu and Yin Yang with equal contribution.}
\affiliation{College of Physics, Mechanical and Electrical Engineering, Jishou University, Jishou 416000, P.R.China}
\affiliation{Guizhou Aerospace Tianma Electromechanical S\&T Co.,LTD, Zunyi 563100, P.R.China}

\author{Chun--Wang Wu}
\affiliation{Department of Physics, National University of Defense Technology, Changsha 410073, P.R.China}
\affiliation{Interdisciplinary Center for Quantum Information, National University of Defense Technology, Changsha 410073, P.R.China}

\author{Wei Wu}
\affiliation{Department of Physics, National University of Defense Technology, Changsha 410073, P.R.China}
\affiliation{Interdisciplinary Center for Quantum Information, National University of Defense Technology, Changsha 410073, P.R.China}

\author{Ping--Xing Chen}
\affiliation{Department of Physics, National University of Defense Technology, Changsha 410073, P.R.China}
\affiliation{Interdisciplinary Center for Quantum Information, National University of Defense Technology, Changsha 410073, P.R.China}

\begin{abstract}
A binary mapping from Fock space of bosonic state to qubits is given. Based on the binary mapping, we construte an algorithm of qubitization of bosons with complexity O(log(N)). As an example, the algorithm of qubitization of bosons in matrix product state to simulate real time dynamics of Yukawa coupling is realized. The calculation error bar is estimated by random sampling method. This proposal may be achieved in superconductivity noisy intermediate--scale quantum computer not far future.
\end{abstract}
\maketitle

\section{Introduction}
\label{intro}
Understanding the space of quantum particle resides, subdividing the space by symmetry and spanning the space by qubits in an economic way are one method of constructing digital quantum simulation algorithm. 
Digital quantum simulation can simulate almost any quantum multi--body problem \cite{feynman1999simulating,wen2007quantum,peskin2018introduction,li2020digital} efficiently, no matter the quantum multi--body systems are strong correlated, ultra--strong coupling or non--linear. 
Quantum multi--body problems are the chain, surface or three--dimensional lattice models interacted by bosons, fermions and anyons located in lattice sites. 
Quantum multi--body problems are important for quantum material (quantum spin chain, quantum spin liquid, quantum topological insulator), biomolecular pharmacy and high energy physics (when quantum field theory being latticed). 
The quantum multi--body system, in second quantized version, has a state vector in a huge Hilbert space transcendent position and time. The state vector is driven by Sch$\ddot{o}$rdinger equation and evolved unitarily with Hamiltonian in exponential.
The transcendence of state vector is the origin of non--local entanglements and correlations of particles.
Each fermion, boson and anyon is operator valued field and spanned by annihilation and creation operators.
The annihilation and creation operators can be written in momentum space, position space or anyother complete space.
The position space in quantum multi--body problem is not fundamental, the annihilation and creation operator be written in position space alwaysly because the position space feeds us stubborn intuition, 
and more importantly for the mathematicians and physicists, the position space always has good symmetry for quantum multi--body problems. 
Even though quantum multi--body system is on lattice, the huge Hilbert space is continuously and governed by symmetry, then the topological analyzes are useful and topological phenomena are important for quantum multi-body problems.

There are underlying algorithm of digital quantum simulation of fermions and bosons.
For fermions, there are Jordan--Wigner \cite{Jordan:1928wi,Batista:2000bx} and Bravyi--Kitaev \cite{bravyi2002fermionic,tranter2015b} transformation  with complexities $O(N)$ and $O(log(N))$, respectively, where $N$ is the number of qubits.
For bosons, there is one--to--one mapping from creation and anihilation operators of boson to operations of qubits with complexity $O(N)$ \cite{somma2003quantum}.  There are also one-to-one and efficient mapping of Fock states from neutral network \cite{torlai2018latent,carleo2019machine,kaestle2021efficient}. 
In this paper, we show an binary mapping from Fock space of boson to Hilbert space spanned by qubits. A novel quantum simulation algorithm of boson, named ``qubitization of bosons'', with complexity $O(log(N))$ is derived from the binary mapping. After the analog version \cite{Casanova2011Quantum,Xiang2018Experimental}, an example of using the algorithm of qubitization of bosons to digitally simulate a quantum many--body problem, non--pertubatively real time dynamics of Yukawa coupling, in matrix product state (MPS) \cite{Vidal2003,Vidal:2003lvx,Schollwock2011} is shown and this proposal may achieved in noisy intermediate--scale quantum computer (NISQ) in near future. 

\section{One--to--one maping from Bosons Fock space to qubits}
The one--to--one mapping of boson Fock space to qubits is \cite{somma2003quantum}
\begin{eqnarray*}
|0\rangle_{x}   &\leftrightarrow& |\uparrow_{0}\downarrow_{1}\downarrow_{2}\cdots \downarrow_{N}\rangle_{x},\\
|1\rangle_{x}   &\leftrightarrow& |\downarrow_{0}\uparrow_{1}\downarrow_{2}\cdots \downarrow_{N}\rangle_{x},\\
|2\rangle_{x}   &\leftrightarrow& |\downarrow_{0}\downarrow_{1}\uparrow_{2}\cdots \downarrow_{N}\rangle_{x},\\
&&\cdots\\
|N\rangle_{x}   &\leftrightarrow& |\downarrow_{0}\downarrow_{1}\downarrow_{2}\cdots \uparrow_{N}\rangle_{x},
\end{eqnarray*}
where $|\downarrow\rangle$ and $|\uparrow\rangle$ are spin down and up qubits.
Then, the truncated boson creation operator is written by Pauli matrices
\begin{eqnarray}
\hat{a}_{x}^{\dagger}=\left(\sum_{i=0}^{N-1}\sqrt{i+1} \sigma^{i}_{-}\sigma^{i+1}_{+}\right)_{x}.
\end{eqnarray}
The particles number operator in this one-to-one mapping is
\begin{eqnarray}
\hat{n}_{x}=\left(\sum_{i=0}^{N} n\frac{\sigma_{z}^{i}+1}{2}  \right)_{x}.
\end{eqnarray}
This boson quantum computation algorithm has complexity O(N), where $N$ is number of qubits.

\section{Binary mapping from bosons Fock space to qubits}
We span the truncated $2^{t}-1$ dimension Fock space of boson in position $x$ by $t$ qubits in binary mapping
\begin{eqnarray}\label{phioc}   \nonumber
|0\rangle_{x}&=&|\uparrow_{1}, \uparrow_{2},...,\uparrow_{t-1},\uparrow_{t}\rangle_{x},\\   \nonumber
|1\rangle_{x}&=&|\uparrow_{1}, \uparrow_{2},...,\uparrow_{t-1},\downarrow_{t}\rangle_{x},\\   \nonumber
|2\rangle_{x}&=&|\uparrow_{1}, \uparrow_{2},...,\downarrow_{t-1},\uparrow_{t}\rangle_{x},\\  \nonumber
|3\rangle_{x}&=&|\uparrow_{1}, \uparrow_{2},...,\downarrow_{t-1},\downarrow_{t}\rangle_{x},\\  \nonumber
&&\cdots \\ 
|2^t-1\rangle_{x}&=&|\downarrow_{1}, \downarrow_{2},...,\downarrow_{t-1},\downarrow_{t}\rangle_{x},
\end{eqnarray}
where $|i\rangle_{x}(i=0,1,\cdots,2^{t}-1)$ is the basis of $i$ bosons occupation space in position $x$.
    The maximal occupation number of boson in position $x$ should be infinite in reality and we truncate it to $2^{t}-1$ in this paper for quantum  simulations. The matrix formulation of $t$ qubits truncated creation operator of boson field is 
\begin{eqnarray}\label{atdagger}
\hat{a}^{\dagger}_{t,x}=\left(\begin{array}{cccccc}
0&0&0&0& &0\\
1&0&0&0& &0\\
0&\sqrt{2}&0&0&\cdots &0\\ 
0&0&\sqrt{3}&0&  &0\\
&&\vdots &&    \ddots  &0\\
0&0&0&0& \sqrt{2^{t}-1}&0
\end{array}\right)_{x},
\end{eqnarray}
where $t$ means we need $t$ qubits to simulate a boson quantum state (truncation number is $2^{t}-1$).
If we write the creation operator of boson as $2^{t}-1$ terms
\begin{eqnarray}   \label{hata}
\hat{a}^{\dagger}_{t,x}=(\sum_{i=1}^{2^t-1}\sqrt{i}\hat{c}_{i}^{(1,t)})_{x} ,
\end{eqnarray}
where $(1,t)$ indices upon $\hat{c}^{(1,t)}_{i}$ means the $\hat{c}^{(1,t)}_{i}$ operator acting on qubits $1$ to $t$ at site $x$. Equation (\ref{atdagger}) tells us that for $t$ equals $1$,
\begin{eqnarray}\label{a1dagger}
\hat{a}^{\dagger}_{1,x}=(\sigma_{-}^{1})_{x}.
\end{eqnarray}
We find a recurrence relation to derive the $2^{t}-1$ truncation creation operator $\hat{a}_{t,x}^{\dagger}$ to $2^{(t+1)}-1$ truncation creation operator $\hat{a}_{t+1,x}^{\dagger}$ 
\begin{eqnarray}\nonumber \label{hatap}
\hat{a}_{t+1,x}^{\dagger}=\left(\sum_{i=1}^{2^t-1}\sqrt{i}I_{+}^{1}\otimes \hat{c}_{i}^{(2,t+1)}+\sqrt{2^t}\sigma_{-}^{1}\otimes \sigma_{+}^{2} ...\otimes \sigma_{+}^{t+1}\right.\\
\left.+\sum_{i=2^t+1}^{2^{t+1}-1}\sqrt{i}I_{-}^{1}\otimes \hat{c}_{i}^{(2,t+1)}\right)_{x},\  \ \  \
\end{eqnarray}
where $\left(\hat{c}^{(2,t+1)}_{i}\right)_{x}$ acting on qubits from $2$ to $t+1$ at position $x$. The definition of $\sigma_{+}, \sigma_{-},I_{+}$ and $I_{-}$ are listed 
\begin{eqnarray}
\sigma_{+}=\frac{1}{2}(\sigma_{x}+i\sigma_{y}), &\quad& \sigma_{-}=\frac{1}{2}(\sigma_{x}-i\sigma_{y}), \\
I_{+}=\frac{1}{2}(I+\sigma_{z}), &\quad& I_{-}=\frac{1}{2}(I-\sigma_{z}).
\end{eqnarray}
The Pauli matrix formulation of any $2^{t}-1$ truncation boson creation operator can be derived from (\ref{hata}), (\ref{a1dagger}) and (\ref{hatap}) and we show several examples as follows.

If we choose $2$ qubits to span the truncated boson Fock space, then we have
\begin{eqnarray}\label{a32} 
\hat{a}_{2,x}^{\dagger}&=&\left(I^{1}_{+}\otimes \sigma_{-}^{2}+\sqrt{2} \sigma_{-}^{1}\otimes \sigma_{+}^{2}+\sqrt{3} I^{1}_{-}\otimes \sigma_{-}^{2}\right)_{x},
\end{eqnarray}
where operations $\sigma^{1}$ and $\sigma^{2}$ acting on ``boson 1'' and ``boson 2'' qubits, respectively,  on position $x$. For the boson creation operator, when qubits number $t$ equals $3$, we have 
\begin{eqnarray*}
\hat{a}_{3,x}^{\dagger}=\left[I^{1}_{+}\otimes I^{2}_{+}\otimes \sigma_{-}^{3}+\sqrt{2}I^{1}_{+}\otimes \sigma_{-}^{2}\otimes \sigma_{+}^{3}+\sqrt{3}I^{1}_{+}\otimes I^{2}_{-}\otimes \sigma_{-}^{3} \right. \\
+\sqrt{4}\sigma_{-}^{1}\otimes \sigma_{+}^{2}\otimes \sigma_{+}^{3}+\sqrt{5}I^{1}_{-}\otimes I^{2}_{+}\otimes \sigma_{-}^{3}\\
\left.+\sqrt{6}I^{1}_{-}\otimes \sigma_{-}^{2}\otimes \sigma_{+}^{3}+\sqrt{7}I^{1}_{-}\otimes I^{2}_{-}\otimes \sigma_{-}^{3}\right]_{x}.
\end{eqnarray*}
The bosons creation operator in position $x$ with qubits number $t$ equal $3$ is represented by Pauli matrices as follow  
\begin{eqnarray*}
\hat{a}_{3,x}^{\dagger}= \frac{1}{8}\left[ \left((1+\sqrt{3}+\sqrt{5}+\sqrt{7}) \sigma^{3}_{x}+2\sqrt{2+\sqrt{3}}  \sigma^{2}_{x}\otimes \sigma^{3}_{x} \right.\right. \\
  + 2 \sqrt{2+\sqrt{3}}  \sigma^{2}_{y}\otimes \sigma^{3}_{y} +2 \sigma^{1}_{x}\otimes \sigma^{2}_{x} \otimes \sigma^{3}_{x}  -2 \sigma^{1}_{x}\otimes \sigma^{2}_{y} \otimes \sigma^{3}_{y} \\
+(1+\sqrt{3}-\sqrt{5}-\sqrt{7}) \sigma^{1}_{z}\otimes  \sigma^{3}_{x}  
+(\sqrt{2}-\sqrt{6}) \sigma^{1}_{z} \otimes \sigma^{2}_{x} \otimes \sigma^{3}_{x}   \\
+(1-\sqrt{3}-\sqrt{5}+\sqrt{7})\sigma^{1}_{z}\otimes \sigma^{2}_{z} \otimes \sigma^{3}_{x}+2 \sigma^{1}_{y}\otimes \sigma^{2}_{x} \otimes \sigma^{3}_{y}\\
  +(1-\sqrt{3}+\sqrt{5}-\sqrt{7}) \sigma^{2}_{z}\otimes \sigma^{3}_{x} + (\sqrt{2}-\sqrt{6}) \sigma^{1}_{z}\otimes \sigma^{2}_{y} \otimes  \sigma^{3}_{y}\\
\left.+2 \sigma^{1}_{y}\otimes \sigma^{2}_{y} \otimes \sigma^{3}_{x}\right) 
+ i \left(2 \sqrt{2+\sqrt{3}}  \sigma^{2}_{x}\otimes \sigma^{3}_{y} - 2 \sqrt{2+\sqrt{3}}  \sigma^{2}_{y}\otimes \sigma^{3}_{x} \right. \\
+(-1+\sqrt{3}-\sqrt{5}+\sqrt{7})  \sigma^{2}_{z}\otimes \sigma^{3}_{y}+2\sigma^{1}_{x}\otimes \sigma^{2}_{x} \otimes \sigma^{3}_{y}  \\
 +2 \sigma^{1}_{x}\otimes \sigma^{2}_{y} \otimes \sigma^{3}_{x} -2 \sigma^{1}_{y}\otimes \sigma^{2}_{x}\otimes \sigma^{3}_{x}+2 \sigma^{1}_{y}\otimes \sigma^{2}_{y} \otimes \sigma^{3}_{y}\\
+ (\sqrt{2}-\sqrt{6}) \sigma^{1}_{z}\otimes \sigma^{2}_{x} \otimes  \sigma^{3}_{y}  -  (\sqrt{2}-\sqrt{6}) \sigma^{1}_{z}\otimes \sigma^{2}_{y} \otimes  \sigma^{3}_{x}\\
- (1+\sqrt{3}+\sqrt{5}+\sqrt{7})  \sigma^{3}_{y}- (1+\sqrt{3}-\sqrt{5}-\sqrt{7}) \sigma^{1}_{z}\otimes \sigma^{3}_{y}\\
\left.\left.-(1-\sqrt{3}-\sqrt{5}+\sqrt{7})\sigma^{1}_{z}\otimes \sigma^{2}_{z} \otimes \sigma^{3}_{y}\right)\right]_{x},
\end{eqnarray*}
and the particle number operator
\begin{eqnarray*}
 \hat{n}_{3,x}=\hat{a}^{\dagger}_{3,x}\hat{a}_{3,x}=\frac{1}{2}(7I-4\sigma_{z}^{1}-2\sigma_{z}^{2}-\sigma_{z}^{3})_{x}.
\end{eqnarray*}
Something usually appearing blocks in quantum simulation of bosons are
\begin{eqnarray*}
{ \hat{n}_{3,x} \hat{n}_{3,x}}=\frac{1}{2}(35I-28\sigma_{z}^{1}-14\sigma_{z}^{2}-7\sigma_{z}^{3}+8\sigma_{z}^{1}\sigma_{z}^{2}\\
+4\sigma_{z}^{1}\sigma_{z}^{3}+2\sigma_{z}^{2}\sigma_{z}^{3})_{x},
\end{eqnarray*}
and
\begin{eqnarray*}
{\hat{a}^{\dagger}_{3,x}\hat{a}^{\dagger}_{3,x}+\hat{a}_{3,x}\hat{a}_{3,x}} =  \frac{1}{4}[(\sqrt{2}+\sqrt{6}+\sqrt{30}+\sqrt{42})\sigma_{x}^{2}\\
+(\sqrt{12}-\sqrt{20})\sigma_{x}^{1}\sigma_{x}^{2}\sigma_{z}^{3}
+(\sqrt{12}-\sqrt{20})\sigma_{y}^{1}\sigma_{y}^{2}\sigma_{z}^{3}\\
+(\sqrt{12}+\sqrt{20})\sigma_{x}^{1}\sigma_{x}^{2}+(\sqrt{2}-\sqrt{6}-\sqrt{30}+\sqrt{42})\sigma_{z}^{1}\sigma_{x}^{2}\sigma_{z}^{3}\\
+(\sqrt{12}+\sqrt{20})\sigma_{y}^{1}\sigma_{y}^{2}+(\sqrt{2}-\sqrt{6}+\sqrt{30}-\sqrt{42})\sigma_{x}^{2}\sigma_{z}^{3}\\
+(\sqrt{2}+\sqrt{6}-\sqrt{30}-\sqrt{42})\sigma_{z}^{1}\sigma_{x}^{2}]_{x}.
\end{eqnarray*}
For qubits number $t$ equals $4$, the Pauli matrices formulation of boson creation operator in position $x$ can be derived from recurrence relation (\ref{hatap})
\begin{eqnarray*}
\hat{a}_{4,x}^{\dagger}=\left[I^{1}_{+}\otimes I^{2}_{+}\otimes  I^{3}_{+}\otimes \sigma_{-}^{4}
+\sqrt{2}I^{1}_{+}\otimes I^{2}_{+} \otimes \sigma_{-}^{3}\otimes \sigma_{+}^{4}\right.\\
+\sqrt{3}I^{1}_{+}\otimes I^{2}_{+}\otimes I^{3}_{-}\otimes \sigma_{-}^{4}+\sqrt{4} I^{1}_{+} \otimes \sigma_{-}^{2}\otimes \sigma_{+}^{3}\otimes \sigma_{+}^{4}\\
 +\sqrt{5} I^{1}_{+} \otimes I^{2}_{-} \otimes I^{3}_{+} \otimes \sigma_{-}^{4}+\sqrt{6} I^{1}_{+}\otimes I^{2}_{-}\otimes \sigma_{-}^{3} \otimes \sigma_{+}^{4}\\
+{ \sqrt{7} I_{+}^{1}\otimes I^{2}_{-}\otimes I_{-}^{3}\otimes \sigma_{-}^{4} }      +\sqrt{8} \sigma_{-}^{1}\otimes \sigma_{+}^{2}\otimes \sigma_{+}^{3}\otimes \sigma_{+}^{4}\\
 +\sqrt{9}I^{1}_{-}\otimes I^{2}_{+}\otimes  I^{3}_{+}\otimes \sigma_{-}^{4}+\sqrt{10} I^{1}_{-}\otimes I^{2}_{+} \otimes \sigma_{-}^{3}\otimes \sigma_{+}^{4}
\\
+\sqrt{11}I^{1}_{-}\otimes I^{2}_{+}\otimes I^{3}_{-}\otimes \sigma_{-}^{4}+\sqrt{12} {I^{1}_{-} \otimes \sigma_{-}^{2}\otimes \sigma_{+}^{3}\otimes \sigma_{+}^{4}} \\
+\sqrt{13} I^{1}_{-} \otimes I^{2}_{-} \otimes I^{3}_{+} \otimes \sigma_{-}^{4} 
+\sqrt{14} I^{1}_{-}\otimes I^{2}_{-}\otimes \sigma_{-}^{3} \otimes \sigma_{+}^{4}\\
\left.+\sqrt{15} {I_{-}^{1}\otimes I^{2}_{-}\otimes \sigma_{-}^{3}\otimes \sigma_{-}^{4}}\right]_{x},
\end{eqnarray*}
and for qubits number $t$ is equal to $5$
\begin{widetext}
\begin{eqnarray*}
\hat{a}_{5,x}^{\dagger}=\left[I^{1}_{+}\otimes I^{2}_{+}\otimes I^{3}_{+}\otimes  I^{4}_{+}\otimes \sigma_{-}^{5}
+\sqrt{2}I^{1}_{+}\otimes I^{2}_{+}\otimes I^{3}_{+} \otimes \sigma_{-}^{4}\otimes \sigma_{+}^{5}+\sqrt{3}I^{1}_{+}\otimes I^{2}_{+}\otimes I^{3}_{+}\otimes I^{4}_{-}\otimes \sigma_{-}^{5}\right.\\
+\sqrt{4} I^{1}_{+}\otimes I^{2}_{+} \otimes \sigma_{-}^{3}\otimes \sigma_{+}^{4}\otimes \sigma_{+}^{5} 
+\sqrt{5} I^{1}_{+}\otimes I^{2}_{+} \otimes I^{3}_{-} \otimes I^{4}_{+} \otimes \sigma_{-}^{5} +\sqrt{6} I^{1}_{+}\otimes I^{2}_{+}\otimes I^{3}_{-}\otimes \sigma_{-}^{4} \otimes \sigma_{+}^{5}\\
+\sqrt{7} {I^{1}_{+}\otimes I_{+}^{2}\otimes I^{3}_{-}\otimes I_{-}^{4}\otimes \sigma_{-}^{5}}+\sqrt{8} I^{1}_{+}\otimes \sigma_{-}^{2}\otimes \sigma_{+}^{3}\otimes \sigma_{+}^{4}\otimes \sigma_{+}^{5}
+\sqrt{9}I^{1}_{+}\otimes I^{2}_{-}\otimes I^{3}_{+}\otimes  I^{4}_{+}\otimes \sigma_{-}^{5}\\
+\sqrt{10} I^{1}_{+}\otimes I^{2}_{-}\otimes I^{3}_{+} \otimes \sigma_{-}^{4}\otimes \sigma_{+}^{5}
+\sqrt{11} I^{1}_{+}\otimes I^{2}_{-}\otimes I^{3}_{+}\otimes I^{4}_{-}\otimes \sigma_{-}^{5}+\sqrt{12} {I^{1}_{+}\otimes I^{2}_{-} \otimes \sigma_{-}^{3}\otimes \sigma_{+}^{4}\otimes \sigma_{+}^{5}}\\
+\sqrt{13}I^{1}_{+}\otimes I^{2}_{-} \otimes I^{3}_{-} \otimes I^{4}_{+} \otimes \sigma_{-}^{5} +\sqrt{14} I^{1}_{+}\otimes I^{2}_{-}\otimes I^{3}_{-}\otimes \sigma_{-}^{4} \otimes \sigma_{+}^{5}
+\sqrt{15} {I^{1}_{+}\otimes I_{-}^{2}\otimes I^{3}_{-}\otimes \sigma_{-}^{4}\otimes \sigma_{-}^{5}}\\
+\sqrt{16}\sigma_{-}^{1}\otimes \sigma_{+}^{2}\otimes \sigma_{+}^{3}\otimes \sigma_{+}^{4}\otimes \sigma_{+}^{5}\sqrt{17}I^{1}_{-}\otimes I^{2}_{+}\otimes I^{3}_{+}\otimes  I^{4}_{+}\otimes \sigma_{-}^{5}
+\sqrt{18}I^{1}_{-}\otimes I^{2}_{+}\otimes I^{3}_{+} \otimes \sigma_{-}^{4}\otimes \sigma_{+}^{5}\\
+\sqrt{19}I^{1}_{-}\otimes I^{2}_{+}\otimes I^{3}_{+}\otimes I^{4}_{-}\otimes \sigma_{-}^{5}
+\sqrt{20} I^{1}_{-}\otimes I^{2}_{+} \otimes \sigma_{-}^{3}\otimes \sigma_{+}^{4}\otimes \sigma_{+}^{5}+\sqrt{21} I^{1}_{-}\otimes I^{2}_{+} \otimes I^{3}_{-} \otimes I^{4}_{+} \otimes \sigma_{-}^{5} \\
+\sqrt{22} I^{1}_{-}\otimes I^{2}_{+}\otimes I^{3}_{-}\otimes \sigma_{-}^{4} \otimes \sigma_{+}^{5}+\sqrt{23} {I^{1}_{-}\otimes I_{+}^{2}\otimes I^{3}_{-}\otimes I_{-}^{4}\otimes \sigma_{-}^{5}}
+\sqrt{24} I^{1}_{-}\otimes \sigma_{-}^{2}\otimes \sigma_{+}^{3}\otimes \sigma_{+}^{4}\otimes \sigma_{+}^{5}\\
+\sqrt{25}I^{1}_{-}\otimes I^{2}_{-}\otimes I^{3}_{+}\otimes  I^{4}_{+}\otimes \sigma_{-}^{5}
+\sqrt{26} I^{1}_{-}\otimes I^{2}_{-}\otimes I^{3}_{+} \otimes \sigma_{-}^{4}\otimes \sigma_{+}^{5}+\sqrt{27} I^{1}_{-}\otimes I^{2}_{-}\otimes I^{3}_{+}\otimes I^{4}_{-}\otimes \sigma_{-}^{5}
\\
+\sqrt{28} {I^{1}_{-}\otimes I^{2}_{-} \otimes \sigma_{-}^{3}\otimes \sigma_{+}^{4}\otimes \sigma_{+}^{5}}+\sqrt{29}I^{1}_{-}\otimes I^{2}_{-} \otimes I^{3}_{-} \otimes I^{4}_{+} \otimes \sigma_{-}^{5} 
+\sqrt{30} I^{1}_{-}\otimes I^{2}_{-}\otimes I^{3}_{-}\otimes \sigma_{-}^{4} \otimes \sigma_{+}^{5}\\
\left.+\sqrt{31} {I^{1}_{-}\otimes I_{-}^{2}\otimes I^{3}_{-}\otimes \sigma_{-}^{4}\otimes \sigma_{-}^{5}}\right]_{x}.
\end{eqnarray*}
\end{widetext}
\section{An example: using the algorithm of qubitization of bosons to digitally simulate Yukawa coupling }
The discrete Hamiltonian in interaction picture of Yukawa coupling is ($\hbar=c=1$)
\begin{eqnarray}\label{L01}
H_{I}=g \sum_{x} \kappa \psi^{\dagger}(x)\psi(x) \phi(x),
\end{eqnarray}
where $x$ is one--dimension position space with lattice spacing $\kappa$. The fermion and scalar fields are represented by creation and annihilation operators \cite{kuypers,nason1984lattice}
\begin{eqnarray}\label{psi}
\phi(x)&=&\frac{1}{\sqrt{2\omega_{0}}}\left(\hat{a}_{t,x}e^{-i\omega_{0}t}+\hat{a}^{\dagger}_{t,x}e^{i\omega_{0}t}\right), \\ 
\psi(x)&=&\frac{1}{\sqrt{2\omega}}\left( \hat{b}_{x}e^{-i\omega t}+ \hat{d}^{\dagger}_{x}e^{i\omega t}\right),
\end{eqnarray}
where $\hat{a}_{t,x}, \hat{b}_{x}$ and $d_{x}$ ($\hat{a}^{\dagger}_{t,x}, \hat{b}^{\dagger}_{x}$ and $d^{\dagger}_{x}$) are annihilation (creation) operators of boson, fermion and anti-fermion. $\omega_{0}$ and $\omega$ are masses of scalar and fermions. The operator $\hat{a}_{t,x}$ means there are $t$ qubits to span the boson truncated Fock space in position $x$. The creation and annihilation operators in position space are Fourier transform version of creation and annihilation operators in momentum space
\begin{eqnarray}
\hat{a}_{t,x}&=&\int \frac{dp}{2\pi}\hat{a}_{t,p}e^{ipx} , \\ 
\hat{b}_{x}=\int \frac{dp}{2\pi}\hat{b}_{p}e^{ipx},  &\ \ & \hat{d}^{\dagger}_{x}=\int \frac{dp}{2\pi}\hat{d}^{\dagger}_{p}e^{-ipx}.
\end{eqnarray}
Then the Hamiltonian is written
\begin{eqnarray}\nonumber \label{Hamiltonian}
H_{I}=\frac{g\kappa}{2\omega\sqrt{2\omega_{0}}} \sum_{x} \left[(\hat{b}^{\dagger}_{x} \hat{b}_{x}+\hat{b}^{\dagger}_{x} \hat{d}^{\dagger}_{x}e^{2i\omega t}
+\hat{d}_{x} \hat{b}_{x}e^{-2i\omega t}\right.\\
\left.+\hat{d}_{x} \hat{d}^{\dagger}_{x})\hat{a}_{t,x}e^{-i\omega_{0}t}
+H.c.\right].\ \ \ \ \ \ 
\end{eqnarray}
The Fock spaces of fermion and anti--fermion are spanned by qubits as follows
\begin{eqnarray}
|0\rangle_{x,s}=| \downarrow \rangle_{x,s},   \quad  |1\rangle_{x,s}= | \uparrow \rangle_{x,s},
\end{eqnarray}
 where $s=1,2=N,P$ is index of anti--fermion and fermion.  The Jordan--Wigner mapping gives us a Pauli matrices representation of the creation operators of fermion and anti--fermion
\begin{eqnarray} \label{bdagger}
\hat{b}^{\dagger}_{x}=-\sigma_{z}^{x,N}\sigma^{x,P}_{+},\quad   \hat{d}^{\dagger}_{x}=  \sigma^{x,N}_{+}, \label{b3}
\end{eqnarray}
where $\sigma^{P}$ and $\sigma^{N}$ acting on qubits of ``fermion'' and ``anti-fermion'', respectively, on position $x$.

The Pauli matrices formulation of Yukawa coupling Hamiltonian in interaction picture from equations (\ref{a32}), (\ref{Hamiltonian}), (\ref{bdagger}) with $t$ equal to $2$ is
\begin{eqnarray}\nonumber
H_{I}&=&\eta \sum_{x} \left[\left( \xi_{1}I+\xi_{2}\sigma^{x,P}_{z}+\xi_{3}\sigma^{x,N}_{z} +\xi_{4}\sigma_{x}^{x,P}\sigma_{x}^{x,N}\right.\right.\\   \nonumber
&&\left.+ \xi_{5}\sigma_{y}^{x,P}\sigma_{y}^{x,N}  + \xi_{6}\sigma_{x}^{x,P}\sigma_{y}^{x,N}+\xi_{7}\sigma_{y}^{x,P}\sigma_{x}^{x,N}    \right)\\   \nonumber
&&\left( \zeta_{1}\sigma_{x}^{2}+\zeta_{2}\sigma_{z}^{1}\sigma_{x}^{2}+\zeta_{3}\sigma_{x}^{1}\sigma_{x}^{2}+\zeta_{4}\sigma_{x}^{1}\sigma_{y}^{2} + \zeta_{5}\sigma_{y}^{2}\right.\\
&&\left.\left.+\zeta_{6}\sigma_{z}^{1}\sigma_{y}^{2}+\zeta_{7}\sigma_{y}^{1}\sigma_{x}^{2}+\zeta_{8}\sigma_{y}^{1}\sigma_{y}^{2} \right)\right]_{x},
\end{eqnarray}
where
\begin{eqnarray}\label{parameters} \nonumber
\eta=\frac{g\kappa}{8\omega\sqrt{2\omega_{0}}},  \quad   \xi_{1}=2 , \\   \nonumber
 \quad \xi_{2}=1, \quad \xi_{3}= -1, \\   \nonumber
     \xi_{4}=-\cos2\omega t,   \quad   \xi_{5}=\cos2\omega t, \\    \nonumber
 \quad  \xi_{6}= \sin2\omega t ,  \quad  \xi_{7}=\sin2\omega t, \\    \nonumber
 \zeta_{1}=(1+\sqrt{3}) \cos\omega_{0}t , \quad \zeta_{2}=(1-\sqrt{3}) \cos \omega_{0}t, \\    \nonumber
  \quad \zeta_{3}= \sqrt{2}\cos\omega_{0}t ,\quad \zeta_{4}=-\sqrt{2}\sin\omega_{0}t ,\\    \nonumber
 \zeta_{5}=(1+\sqrt{3}) \sin\omega_{0}t , \quad \zeta_{6}= (1-\sqrt{3}) \sin\omega_{0}t, \\    
 \quad \zeta_{7}=\sqrt{2}\sin\omega_{0}t  , \quad \zeta_{8}= \sqrt{2}\cos\omega_{0}t .
\end{eqnarray}
The evolution of quantum state of quantum field theory in interaction picture is driven by the time--evolution operator $U(t,t_{0})$
\begin{eqnarray}
\label{stateevolution}
|\Psi(t)\rangle = U(t,t_{0}) |\Psi(t_{0})\rangle.
\end{eqnarray}
 The $|\Psi(t_0)\rangle$ is initial state and can be set by hand. The time evolution operator $U(t,t_{0})$ satisfy
\begin{equation}
U(t,t_{0})={\underbrace { U(t,t-\Delta t)\cdots  U(t_{0}+\Delta t,t_{0}) }_{n_{t}}},
\end{equation}
where $n_{t}\cdot\Delta t=t-t_{0}$. Eliminating the error bar from trancation of occupation of bosons, the time evolution operator $U(t+\Delta t,t)$ is approximated with
\begin{eqnarray}\label{DeltatU}  \nonumber
U(t+\Delta t,t)=\lim_{N\rightarrow \infty} e^{-i\frac{1}{N}\left[H_{I}(t+\frac{N-1}{N}\Delta t)+H_{I}(t+\frac{N-2}{N}\Delta t)\cdots +H_{I}(t)\right]\Delta t}\\
\approx e^{-iH_{I}(t^{\prime})\Delta t},\ \ \ \ \  \end{eqnarray}
where $t^{\prime}\in [t,t+\Delta t]$. The error bar from nondeterminacy of the $t^{\prime}$ can be estimated by number of random sampling $t^{\prime}$ in time interval $[t,t+\Delta t]$.

\begin{figure}
\begin{center}
\includegraphics[width= 86 mm]{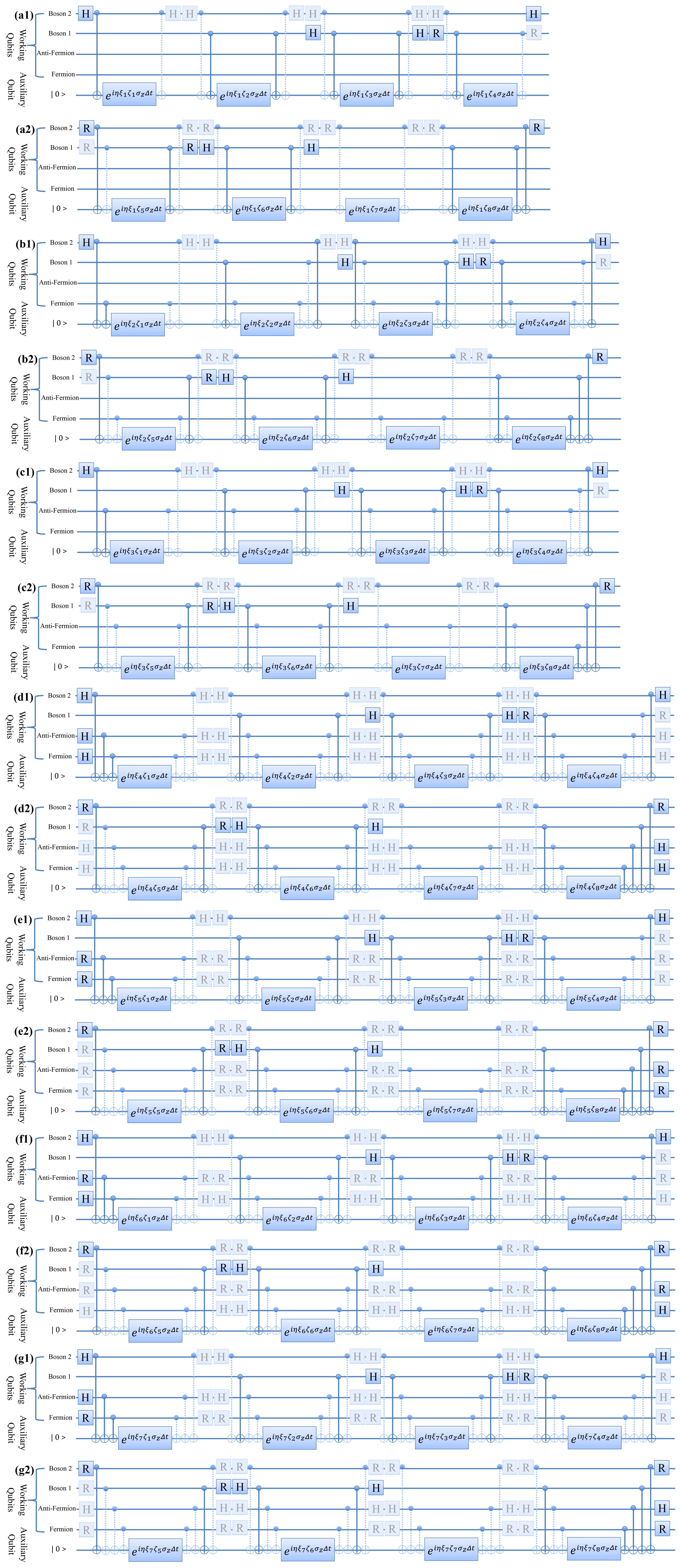}
\caption{\label{pic1} Quantum circuits to simulate time evolution operator (\ref{DeltatU}) is shown. There are lots of single qubit operations $H,R$ and CNOT operations being cancelled and we show them by light blue and dotted lines.}
\end{center}
\end{figure}

The quantum circuit to simulate (\ref{DeltatU}) is shown in Fig.~\ref{pic1} and the parameters in Fig.~\ref{pic1}. The $H$ in Fig.~\ref{pic1} is Hadamard operation and $R$ is
\begin{eqnarray}
R=\frac{1}{\sqrt{2}}\left(\begin{array}{cc} 
1&-i\\
i&-1
\end{array}\right).   
\end{eqnarray}
We measure the particle number operator of boson and fermions to get real time dynamics of occupation probability $\rho(t)$ as follow
\begin{eqnarray}\label{rhoboson}
\rho_{b}(t)&=&\sum_{x}\langle \Psi(x,t)|\hat{n}_{x,b}|\Psi(x,t)\rangle,\\   \label{rhofermion}
\rho_{s,f}(t)
&=&(-1)^{s-1}\sum_{x}\langle \Psi(x,t)|\hat{n}_{x,s}|\Psi(x,t)\rangle,
\end{eqnarray}
where the Pauli matrix formulation of particle number operator of boson is
\begin{eqnarray}
\hat{n}_{x,b}= \hat{a}^{\dagger}_{2,x} \hat{a}_{2,x}=\frac{1}{2}(3I-2\sigma_{z}^{1}-\sigma_{z}^{2})_{x},
\end{eqnarray}
and the particle number operators of fermion and anti--fermion are
\begin{eqnarray}
\hat{n}_{x,P}&=&\hat{b}^{\dagger}_{x} \hat{b}_{x}=\frac{1}{2}(I+\sigma_{z}^{x,P}),\\
 \hat{n}_{x,N}&=&\hat{d}^{\dagger}_{x} \hat{d}_{x}=\frac{1}{2}(I+\sigma_{z}^{x,N}).
\end{eqnarray}
For anti--fermion, we multiplier a ``$-1$'' factor by hand in front of the occupation density in (\ref{rhofermion}). 

\begin{figure}
\begin{center}
\includegraphics[width=86 mm]{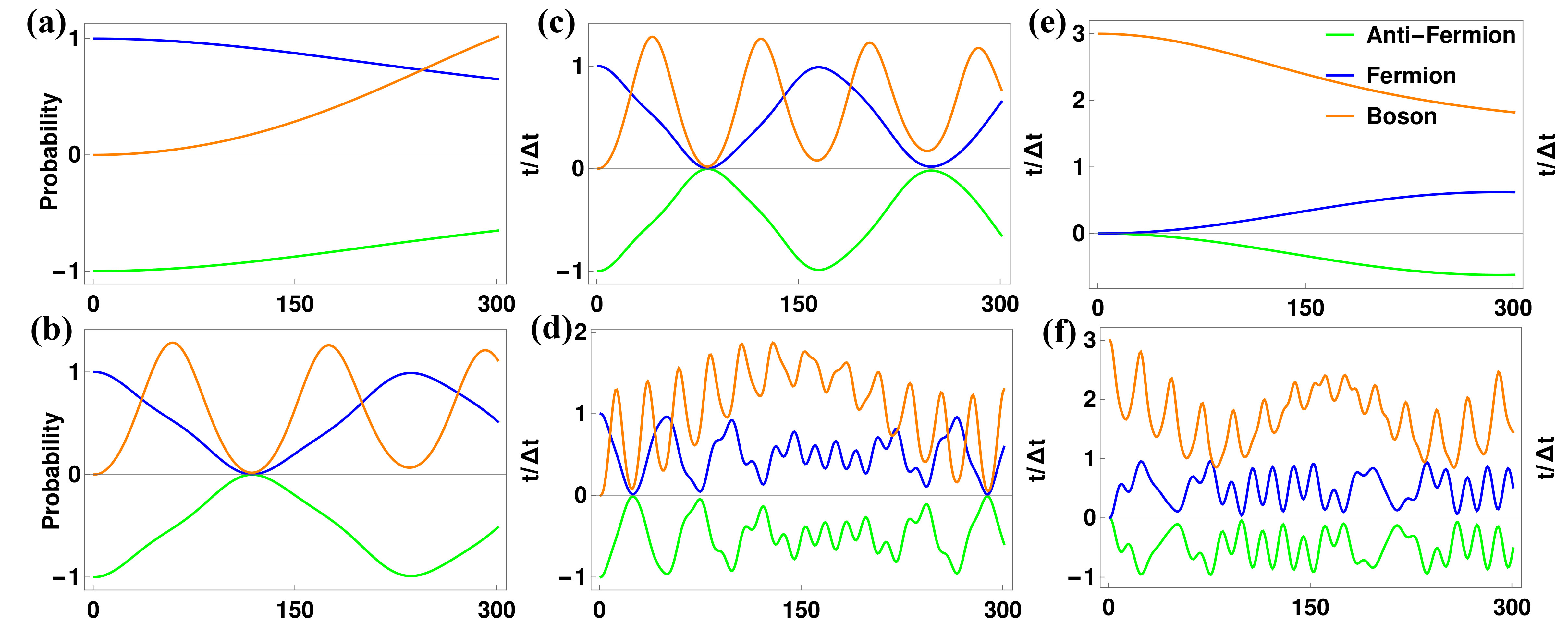}
\caption{\label{pic2} Simulation results with varies initial states and coupling constants calculated by MPS are shown. The initial state of (a), (b), (c) and (d) is a pair of fermion anti-fermion in the site $x$. The initial state of (e) and (f) is 3 bosons on site $x$. The coupling constants of (a), (b), (c), (d), (e) and (f) are 1, 6.95, 10, 34.75, 1 and 34.75, respectively. The horizontal axis is time slice number $l=t/\Delta t$ and the vertical axis is probability for fermion, anti-fermion and boson. The blue, green and yellow lines are real time dynamics of occupation density $\rho(t)$ defined by (\ref{rhoboson}) and (\ref{rhofermion}) of fermion, anti--fermion and boson, respectively. }
\end{center}
\end{figure}

\begin{figure}
\begin{center}
\includegraphics[width= 60 mm]{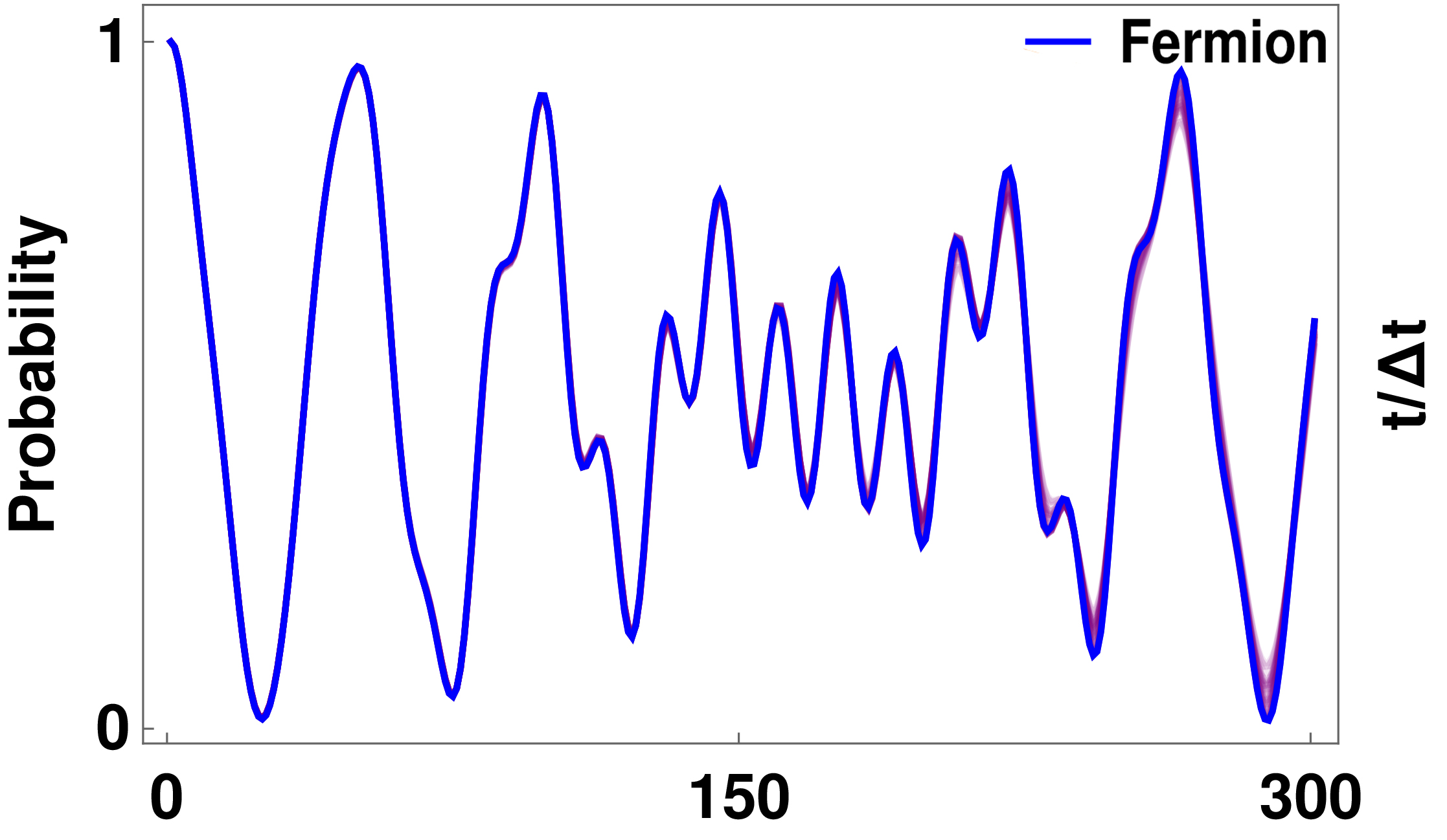}
\caption{\label{errorbar} Simulation results with 50 times random $t^{\prime}$ with initial state of a pair of fermion anti-fermion in the site $x$ and coupling constant 34.75. The blue line is a typical simulation results of probability evolution of fermions. The purple shadow is error bar estimation from 50 times random $t^{\prime}$ sampling.}
\end{center}
\end{figure}

\label{result}
We realize the algorithm to simulate the real time dynamics of Yukawa coupling in MPS and the default parameters are taken to be
\begin{eqnarray*}
\kappa=0.5,\quad \Delta t=0.1, \quad n_{x}=1,\quad n_{t}=300, \\
 t=l\cdot\Delta t, (l=0,\cdots n_{t}),\quad \omega=6.95,\quad \omega_{0}=1,
\end{eqnarray*}
where we set one position site $x$ then $n_{x}$ is equal to $1$. The $n_{t}$ is time slices number, the $\Delta t$ is time step length. The non--perturbative MPS calculation results are shown in Fig.~\ref{pic2}.  The Fig.~\ref{pic2}(a), (b), (c) and (d) are real time dynamics evolutions with fermion, antifermion pair initial state in coupling constant 1, 6.95, 10 and 34.75, respectively. With the increasing of coupling constant, the fermion anti-fermion pair annihilating to bosons quickly. At Fig.~\ref{pic2}(d), the non-linear dynamics of fermion anti--fermion pair and bosons are clearly emerging in ultra-strong coupling region. The initial state of Fig.~\ref{pic2}(e) and (f) is 3 bosons in site $x$ with coupling constant 1 and 34.75. The real time dynamics of Fig.~\ref{pic2}(e) and (f) show that the boson create fermion anti--fermion pairs from vaccum through Yukawa coupling. In ultra--strong coupling Fig.~\ref{pic2}(f), the real time dynamics of fermion anti--fermion pair and bosons are non-linear. To estimate the error of the calculation results, we set the $t^{\prime}$ is a random number in MPS
\begin{eqnarray}
t^{\prime}=t+random(0,1)*\Delta t,
\end{eqnarray}
 and each $t^{\prime}$ sample $50$ times, where $random(0,1)$ gives us a random number from $0$ to $1$. As an example, the error bar of the real time dynamics evolution of fermions with parameter and initial state setting of Fig.~\ref{pic2}(d) is showed by purple shadow behind blue line (see Fig.~\ref{errorbar} ). Fig.~\ref{errorbar} shows that the error bar of the real time dynamics evolution of Fermions in Yukawa coupling is apparent when probability of fermions with drastic change. When coupling constant is small and the dynamics is linear, the error bar shadow almost cannot be seen.

\begin{figure}
\begin{center}
\includegraphics[width= 48 mm]{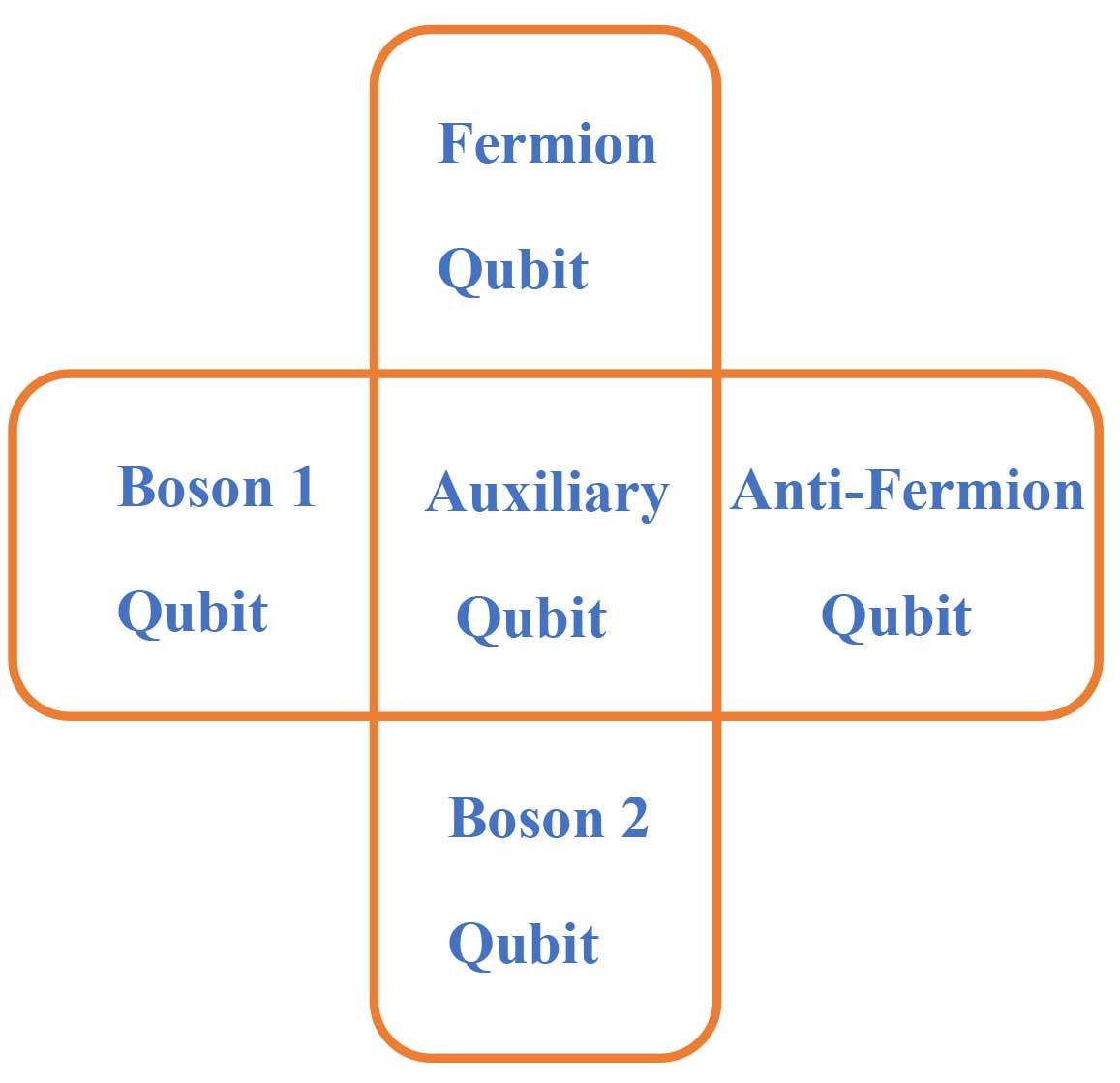}
\caption{\label{pic3} The relations of 5 qubits to digitally simulate Yukawa coupling. The connected qubits need to perform CNOT operations between them.}
\end{center}
\end{figure}

The quantum algorithm of real time dynamics simulation of Yukawa coupling needs 
 $117$ CNOT operations in each $\Delta t$ unitary time evolution operator $U(t+\Delta t, t)$ from Fig.~\ref{pic1}. 
The quantum circuits Fig.~\ref{pic1} can be ran in NISQ computer also. For example, we set the $\Delta t$ equals to $1/4, 1/3, 1/2$ and $n_{t}$ equal to $10$; then for total fidelity $70\%$, the fidelity of each CNOT operation should higher than
\begin{eqnarray}
^{117\times 10}\sqrt{70 \%}=99.9695 \%.
\end{eqnarray}
At present, the average fidelity of CNOT operation in IBMQ\_santiago 5 qubits superconductivity NISQ computer is $99.4395\%$. We hope to see the digital quantum simulation algorithm of Yukawa coupling be ran in superconductivity NISQ computers such as IBMQ not far future.

\section{Summary}
\label{con}
We shown an binary mapping from bosons Fock space to qubits which reduces the complexity of bosons simulation algorithm to $O(log(N))$. As an example, the mapping derives a digital quantum simulation algorithm of Yukawa coupling. We realized the algorithm in MPS and shown the non--perturbative calculation results. It proved that the MPS can work in ultra--strong coupling and non--linear regions of quantum field theory. We demonstrated an experimental realization in superconductivity NISQ computer for the digital quantum simulation of Yukawa coupling.


\begin{acknowledgements}
This work is supported by the National Basic Research Program of China under Grant No. 2016YFA0301903. We thank Ming Zhong, Kihwan Kim for valuable discussions. 
\end{acknowledgements}



\providecommand{\href}[2]{#2}\begingroup\raggedright\endgroup

\end{document}